\documentclass{article}

\usepackage[nonatbib, preprint]{cig}

\usepackage[utf8]{inputenc} 
\usepackage[T1]{fontenc}    
\usepackage{hyperref}       
\usepackage{arydshln}
\usepackage[export]{adjustbox}
\usepackage{tikz}
\usepackage{cite}
\usepackage{booktabs}
\usepackage{multirow}
\usepackage{amssymb}
\usepackage{outlines}
\def\x{{\mathbf x}}

\usepackage{amsmath, amssymb, amsfonts, mathtools, bm}
\usepackage{nicefrac}
\usepackage{bbm}
\usepackage{siunitx}
\usepackage{pifont}
\usepackage{algorithm}
\usepackage{algpseudocode}
\usepackage{amsthm}
\usepackage{calc}
\usepackage{cite}

\usepackage{graphicx}
\usepackage{tabularx, threeparttable, tabularray}
\UseTblrLibrary{booktabs}
\usepackage{booktabs}
\usepackage{multirow}
\usepackage{wrapfig}
\usepackage{arydshln}
\usepackage{hhline}
\usepackage{caption}
\usepackage{diagbox}

\usepackage{microtype}
\usepackage{setspace}
\usepackage{lipsum}
\usepackage{soul}
\usepackage{xcolor}
\definecolor{darkblue}{rgb}{0, 0, 0.85}
\definecolor{lightgreen}{rgb}{.85,1,.85}
\definecolor{lightred}{rgb}{1,.85,.85}
\definecolor{lightblue}{rgb}{.85,.85,1}
\definecolor{pink}{HTML}{EB346F}
\newcommand{\hlgreen}[1]{{\sethlcolor{lightgreen}\hl{#1}}}

\newcommand{\hlblue}[1]{{\sethlcolor{lightblue}\hl{#1}}}

\DeclarePairedDelimiterX{\infdivx}[2]{(}{)}{#1\;\delimsize\|\;#2}

\renewcommand{\vec}[1]{\bm{#1}}

\newcommand{\A}{{\vec{A}}}

\def\x{\vec{x}}  
\def\y{\vec{y}}  
\def\z{\vec{z}}  
\def\e{\vec{e}}  
\def\s{\vec{s}}

\def\argmin{\mathop{\mathsf{arg\,min}}}

\usepackage{pifont}

\hypersetup{linkcolor=black, citecolor=black, filecolor=black, urlcolor=pink}

\definecolor{darkblue}{rgb}{0,0 ,0.542}
\definecolor{lightgreen}{rgb}{.9,1,.9}
\definecolor{lightred}{rgb}{1,.415,.415}
\definecolor{lightblue}{rgb}{.415,.415,1}
\renewcommand{\vec}[1]{\bm{#1}} 

\definecolor{lightgreen}{rgb}{.85,1,.85}
\definecolor{lightred}{rgb}{1,.85,.85}
\definecolor{lightblue}{rgb}{.85,.85,1}
\definecolor{pink}{HTML}{EB346F}

\definecolor{darkblue}{rgb}{0, 0, 0.85}
\definecolor{lightgreen}{rgb}{.85,1,.85}
\definecolor{lightred}{rgb}{1,.85,.85}
\definecolor{lightblue}{rgb}{.85,.85,1}
\definecolor{pink}{HTML}{EB346F}
\hypersetup{
    colorlinks = true,
    citecolor = darkblue,
    linkcolor = {black}
}

\title{Measurement Score-Based MRI Reconstruction \\with Automatic Coil Sensitivity Estimation}

\vspace{5em}
\author{%
\normalsize Tingjun Liu\thanks{Equal contribution.} \quad Chicago Y. Park\footnotemark[1] \quad Yuyang Hu \\[1.0em] Hongyu An \quad Ulugbek S. Kamilov\\[1.25em]
\small \textnormal{WashU, USA}\\[0.5em]
\footnotesize \texttt{\{l.tingjun, chicago, h.yuyang, hongyuan, kamilov\}@wustl.edu}
}

\begin{document}

\maketitle

\begin{abstract}
    Diffusion-based inverse problem solvers (DIS) have recently shown outstanding performance in compressed-sensing parallel MRI reconstruction by combining diffusion priors with physical measurement models. However, they typically rely on pre-calibrated coil sensitivity maps (CSMs) and ground truth images, making them often impractical: CSMs are difficult to estimate accurately under heavy undersampling and ground-truth images are often unavailable.
    We propose Calibration-free Measurement Score-based diffusion Model (C-MSM), a new method that eliminates these dependencies by jointly performing automatic CSM estimation and self-supervised learning of measurement scores directly from k-space data.
    C-MSM reconstructs images by approximating the full posterior distribution through stochastic sampling over partial measurement posterior scores, while simultaneously estimating CSMs.
    Experiments on the multi-coil brain fastMRI dataset show that C-MSM achieves reconstruction performance close to
    DIS with clean diffusion priors --- even without access to clean training data and pre-calibrated CSMs.
\end{abstract}

\section{Introduction}

Magnetic resonance imaging (MRI) reconstruction seeks to recover a high–quality image from undersampled k-space data. In parallel MRI (PMRI), multiple coils acquire measurements modulated by coil sensitivity maps (CSMs). Let \(\x \in \mathbb{C}^{p}\) be the unknown image and \(\y=(\y_1,\ldots,\y_{n_c})\) the multi-coil measurements from \(n_c\) coils. The acquisition model is
\begin{equation}
\y = \A \x + \e,
\label{eq:global-model}
\end{equation}
where \(\A=(\A_1,\ldots,\A_{n_c})\) is the forward operator and \(\e\) is measurement noise. For coil \(k\),
\begin{equation}
\y_k = \vec{S} \vec{F} \vec{C}_k \x + \e_k, \qquad k=1,\ldots,n_c,
\label{eq:coil-model}
\end{equation}
with \(\vec{C}_k\) the CSM, \(\vec{F}\) the discrete Fourier transform, and \(\vec{S}\) the undersampling operator.

\noindent
Reconstruction is often formulated as optimization:
\begin{equation} \label{eq:begin_inv_problem}
    \widehat{\x} \in \argmin_{\x \in \mathbb{C}^n} g(\x) + h(\x),
\end{equation}
where $g(\x)$ is a data-fidelity term quantifying consistency with the observed measurements $\y$ and $h(\x)$ is a regularizer imposing prior knowledge on $\x$.
The formulation in \eqref{eq:begin_inv_problem} corresponds to the maximum a posteriori (MAP) estimator when
\begin{equation}
    g(\x) = -\log \, p(\y|\x) \text{ and } h(\x) = -\log \,p(\x) \;,
\end{equation}
where $p(\y|\x)$ denotes the likelihood that relates  $\x$ to the measurements $\y$, and $p(\x)$ represents the prior distribution.
For linear inverse problems of form \( \y = \vec{A}\x + \vec{e} \), where \( \vec{A} \in \mathbb{C}^{m \times p} \) represents the measurement operator and \( \vec{e} \in \mathbb{C}^m \) is the additive white Gaussian noise (AWGN), the data-fidelity term becomes \( g(\x) = \frac{1}{2} \|\y - \vec{A}\x\|^2_2 \).

\begin{figure*}[t]
    \centering
    \includegraphics[width=1.0\linewidth]{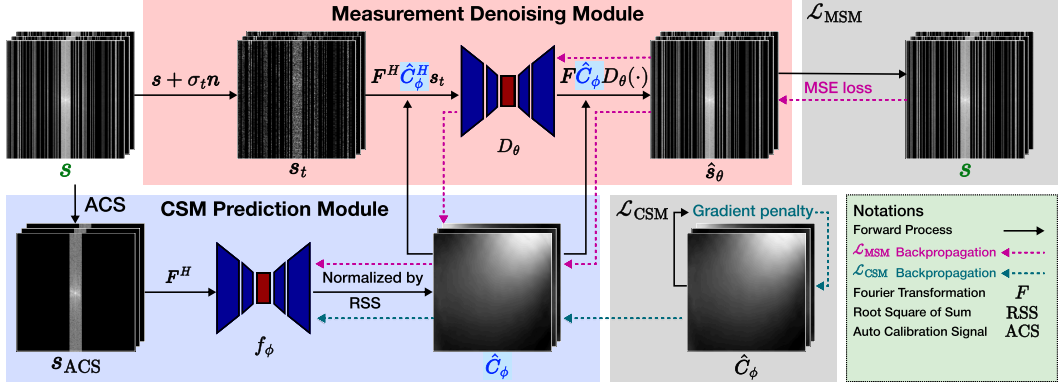}
    \caption{Illustration of the proposed Calibration-free Measurement Score-based diffusion Model (C-MSM) training pipeline, which operates solely on subsampled measurements. C-MSM jointly learns coil sensitivity maps (CSMs) and measurement scores directly from undersampled k-space data.}
    \label{fig:training_pipeline}
\end{figure*}

Recently, diffusion models have emerged as powerful image priors for inverse problems. By learning the distribution of clean images, they provide regularization that can be coupled with measurement consistency to achieve state-of-the-art reconstructions~\cite{daras2024surveydiffusionmodelsinverse,10.1093/bjrai/ubae013}. While early approaches rely on large datasets of high-quality ground truth images, there is growing interest in self-supervised and measurement-driven diffusion methods that eliminate the need for ground truth~\cite{park2025measurementscorebaseddiffusionmodel, aali2025ambient, daras2023ambient}. This direction is relevant for MRI, where acquiring ground truth might be impractical due to long scan times and clinical constraints.
A notable advance in this area is the measurement score-based diffusion model (MSM)~\cite{park2025measurementscorebaseddiffusionmodel}, which enables training diffusion models directly using k-space measurements without access to clean images. MSM methods have demonstrated strong performance both in modeling image priors and in solving inverse problems.

Despite this progress, existing work on self-supervised training of diffusion models has largely overlooked the critical challenge of coil sensitivity estimation. Most diffusion-based inverse problem solvers still rely on pre-estimated CSMs, which are difficult to obtain under heavy undersampling and vary significantly across scans. While some work has considered the joint reconstruction of the image and estimation of coil sensitivities~\cite{Li_Blind_MICCAI2024, Chen2025}, such approaches are trained with fully-sampled reference images, which limits applicability when only undersampled measurements are available.

In this work, we address this gap by proposing \textit{Calibration-free Measurement Score-based diffusion Model (C-MSM)}, a self-supervised diffusion-based framework for PMRI reconstruction that jointly learn predicting CSM estimation and measurement score directly from undersampled k-space data.
During inference, C-MSM approximates the full posterior distribution via stochastic sampling over partial measurement posterior scores, while simultaneously refining CSMs during the diffusion process. 
We validate C-MSM on the multi-coil fastMRI dataset and show that it outperforms self-supervised baselines while approaching the performance of clean prior-based methods.

\section{Background}

\medskip\noindent
\textbf{Diffusion Models for Solving Inverse Problems.} Inverse problems are often solved by combining a data-fidelity term with a prior, as in \eqref{eq:begin_inv_problem}.
Recent progress in generative modeling has led to the use of diffusion models as expressive learned priors for solving inverse problems. These models approximate the score function $\nabla_{\x_t} \log p_{\sigma_t}(\x_t)$ over a wide range of noise levels $\sigma_t$, where $p_{\sigma_t}(\x_t)$ denotes the distribution of data corrupted with Gaussian noise of variance $\sigma_t^2$. Learning scores at multiple noise levels allows the sampling process to begin from pure noise and gradually refine the signal, ultimately recovering a clean and detailed image.

To incorporate measurement consistency, recent methods alternate between denoising with the learned score and enforcing consistency with the forward model. Diffusion posterior sampling (DPS)~\cite{chung2023dps} adjusts the diffusion iterate itself using a gradient step toward the measurement manifold, while denoising diffusion null-space models (DDNM)~\cite{wang2023ddnm} enforce consistency by projecting the denoised estimate onto the measurement constraint. These complementary strategies demonstrate that reconstruction quality depends not only on the prior but also on the design of the data-consistency update.

Building on this progress, a diffusion-based study has also tackled coil-sensitivity estimation in PMRI reconstruction. 
A proximal alternating framework is introduced to jointly estimate MR images and coil sensitivity maps during diffusion sampling, where diffusion models act as learned proximal operators to solve blind parallel MRI reconstruction~\cite{Li_Blind_MICCAI2024}.
However, this approach requires the pretrained diffusion model trained with fully-sampled reference images, which limits applicability when only undersampled measurements are available.

\medskip\noindent
\textbf{Self-Supervised Diffusion-Based MRI Reconstruction.} Self-supervised diffusion models have recently emerged as effective priors for MRI reconstruction, without relying on clean ground-truth data for training. The ambient diffusion model~\cite{daras2023ambient} achieves this by applying an additional subsampling step during training and learns to predict the original subsampled measurement. Ambient diffusion posterior sampling (A-DPS)~\cite{aali2025ambient} extends this framework to inverse problems by combining ambient diffusion~\cite{daras2023ambient} with DPS-style data consistency~\cite{chung2023dps}.
Measurement score-based diffusion models~\cite{park2025measurementscorebaseddiffusionmodel} take a different approach by directly learning measurement score functions in measurement space by adding noise onto subsampled measurements. This enables image generation and solving inverse problems through stochastic prior or posterior sampling, which combine mini-batch measurement scores and measurement posterior scores, respectively.

Existing self-supervised diffusion-based approaches still assume access to accurate pre-calibrated CSMs. This reliance limits their practicality in scenarios where both pre-calibrated CSMs and ground-truth images are unavailable.
Outside the diffusion family, a joint learning approach has been proposed to estimate CSMs and perform reconstruction from paired undersampled/noisy measurements~\cite{hu2024spicerselfsupervisedlearningmri}, but it requires paired acquisitions and does not exploit diffusion-based learning.

\section{Methods}

We propose C-MSM as a self-supervised diffusion framework for MRI reconstruction that jointly estimates CSMs and learns measurement scores directly from undersampled k-space data. C-MSM consists of two components: a CSM prediction network and a measurement score-based diffusion network, which are trained together under a joint objective function.

\medskip\noindent
\textbf{C-MSM Training.}  
We assume C-MSM has access only to partial observations of an unknown fully-sampled multi-coil measurement $\z \in \mathbb{C}^n$, 
\begin{equation}
    \s = \vec{S}\z \in \mathbb{C}^m,
\end{equation}
where $\vec{S}\in \{0, 1\}^{m\times n}$, with $m < n$, is a subsampling mask drawn from distribution $p(\vec{S})$.

The first step in C-MSM is CSM prediction. From the auto-calibration signal (ACS) region $\s_{\text{ACS}}$ of $\s$, we transform the data into the image domain and input it to CSM prediction network $f_\phi$:
\begin{equation}
\hat{\vec{C}}_\phi(\vec{s}_{\text{ACS}}) \;=\; f_\phi\!\left(\vec{F}^{H}\vec{s}_{\text{ACS}}\right).
\end{equation}
The estimated CSMs $\hat{\vec{C}}_\phi$ are then normalized by their root-sum-of-squares (RSS) magnitude to balance the contributions across coils and keep the forward and adjoint operators stable during reconstruction.

Next, for the Measurement Denoising Module, we apply the forward diffusion step by corrupting the measurements with Gaussian noise:
\begin{equation}
\vec{s}_t = \vec{s} + \sigma_t \vec{n}, \qquad \vec{n} \sim \mathcal{N}(0,\vec{I}),
\end{equation}
where $\sigma_t$ is the noise level. To reduce the high dimensionality of multi-coil k-space data, $\vec{s}_t$ is transformed into the image domain, where coil information is combined using the predicted CSMs. The diffusion model $\mathsf{D}_\theta$ is then applied in the image domain, and the result is mapped back to measurement space:
\begin{equation} \label{eq:denoising_s}
\hat{\vec{s}}_{\theta}(\vec{s}_t;\sigma_t, \hat{\vec{C}}_{\phi}) 
= \vec{F}\hat{\vec{C}}_\phi \,\mathsf{D}_\theta\!\left(\vec{F}^{H} \hat{\vec{C}}_\phi^{H} \vec{s}_t;\sigma_t\right).
\end{equation}
The model is trained by minimizing the MSE loss between true and the predicted subsampled measurements:
\begin{equation}
\mathcal{L}_{\text{MSM}}(\theta;\hat{\vec{C}}_\phi)
= \mathbb{E}_{\vec{s}_t,\,t}
\!\left[
  \left\| \vec{s} - \hat{\vec{s}}_{\theta}(\vec{s}_t;\sigma_t, \hat{\vec{C}}_{\phi}) \right\|_2^2
\right].
\end{equation}
While $\mathcal{L}_{\text{MSM}}$ updates both the diffusion network and the CSM predictor through its dependence on $\hat{\vec{C}}_\phi$, this loss alone does not guarantee the known property that coil sensitivity maps are spatially smooth. To explicitly enforce this property, we introduce an additional gradient penalty:
\begin{equation}
\mathcal{L}_{\text{CSM}}(\phi) = \|\nabla \hat{\vec{C}}_\phi(\vec{s}_{\text{ACS}})\|_2^2.
\end{equation}
The total loss is then
\begin{equation}
\mathcal{L}_{\text{total}}(\theta,\phi) 
= \mathcal{L}_{\text{MSM}}(\theta;\hat{\vec{C}}_\phi(\vec{s}_{\text{ACS}})) 
+ \lambda \,\mathcal{L}_{\text{CSM}}(\phi),
\label{eq:total_loss}
\end{equation}
where $\lambda$ is the hyperparameter that balances smoothness and MSM losses.
An overview is shown in Figure~\ref{fig:training_pipeline}.

\medskip\noindent
\textbf{C-MSM Sampling for MRI Reconstruction.}
In conditional sampling, our objective is to reconstruct a fully-sampled measurement given noisy subsampled observations
\begin{equation}
    \y = \vec{H}\z + \e,
\end{equation}
where $\vec{H}\in \mathbb{C}^{m \times n}$ is the sampling operator, $\z \in \mathbb{C}^n$ is the fully-sampled measurement, and $\e \sim \mathcal{N}(\vec{0}, \eta\vec{I})$ is Gaussian noise with noise level $\eta$.

C-MSM predicts CSMs using the pretrained CSM module, which takes as input the ACS region $\vec{y}_{\text{ACS}}$ of $\y$ and computes
\begin{equation}
\hat{\vec{C}}_\phi(\vec{y}_{\text{ACS}}) = f_\phi\!\left(\vec{F}^{H}\vec{y}_{\text{ACS}}\right),
\end{equation}
and normalize the estimated CSMs $\hat{\vec{C}}_\phi(\vec{y}_{\text{ACS}})$ by their RSS magnitude, as in training.

To use the pretrained C-MSM, which was trained to denoise subsampled measurements $\s_t$ toward a fully-sampled diffusion iterate $\z_t$, we draw $w$ random subsampling operators $\vec{S}^{(i)}$, $i \in \{1,\dots,w\}$, and apply them to $\z_t$ as
\begin{equation}
    \s_t^{(i)} = \vec{S}^{(i)}\z_t
\end{equation}
is denoised by \eqref{eq:denoising_s}, producing $\hat{\s}^{(i)}_{\theta}(\s^{(i)}_t; \sigma_t, \hat{\vec{C}}_\phi(\vec{y}_{\text{ACS}}))$.

To enforce data consistency, we update the overlapping coordinates between $\y$ and $\hat{\s}_t^{(i)}$:
\begin{equation}
    \tilde{\s}_{\theta}^{(i)} = \hat{\s}_{\theta}^{(i)} - \gamma_t \nabla \|\y^{(i)} - \tilde{\vec{H}}^{(i)}\hat{\s}_{\theta}^{(i)}\|_2^2,
\end{equation}
where $\tilde{\vec{H}}^{(i)} := \vec{S}^{(i)}\vec{H}^{H}\vec{H}\vec{S}^{(i)H}$ restricts $\vec{H}^{H}\vec{H}$ to coordinates selected by $\vec{S}^{(i)}$.

To combine the $w$ denoised measurements, we follow the weighting sum scheme introduced in~\cite{park2025measurementscorebaseddiffusionmodel}:
\begin{equation}
    \vec{W} := \left[\max\!\left(\sum^w_{i=1} \text{diag}(\vec{S}^{(i)H}\vec{S}^{(i)}), 1\right)\right]^{-1},
\end{equation}
which balances overlapping regions after projection into the fully-sampled space. The elementwise maximum prevents division by zero in regions not covered by any subsampling operator. The fully-sampled estimate is then obtained by weighted averaging:
\begin{equation}
    \tilde{\vec{z}}_{\theta} = \vec{W}\sum^w_{i=1}\vec{S}^{(i)H}\tilde{\s}_{\theta}^{(i)}.
\end{equation}
Finally, the next iterate is sampled given current iterate $\z_t$ and denoised estimate $\tilde{\vec{z}}_\theta$:
\begin{equation}
\vec{z}_{t-1} \sim p\!\left(\vec{z}_{t-1}\mid \z_t, \tilde{\vec{z}}_\theta\right).
\end{equation}

\section{Numerical Evaluations}

\medskip\noindent
\textbf{Experimental Setup.}
We evaluate C-MSM on parallel MRI (PMRI) reconstruction using the multi-coil fastMRI dataset~\cite{zbontar2018fastmri}.
We perform compressed-sensing parallel MRI reconstruction on 80 held-out test images at acceleration factors $\times4$ and $\times8$, and at measurement noise level $\sigma = 0.01$, with 20 ACS lines provided to all methods during testing.

\medskip\noindent
\textbf{Training Setup.}
All models, including baselines, use the same diffusion backbone~\cite{dhariwal2021beat} and are trained from scratch for one million iterations on a single NVIDIA RTX3090 GPU.
For training, we use 2,000 center-cropped T2-weighted slices ($256\times256$, complex-valued).
C-MSM is trained solely from subsampled measurements, without access to fully-sampled ground-truth images and pre-calibrated CSMs, and we specified the smoothness parameter $\lambda$ in \eqref{eq:total_loss} with $\lambda=1000$.
For comparison, MSM~\cite{park2025measurementscorebaseddiffusionmodel} and A-DPS~\cite{aali2025ambient} is trained with pre-calibrated CSMs but no fully-sampled images, while a second baseline diffusion prior is trained under the optimal setting with both ground-truth images and pre-calibrated CSMs.

\begin{table*}[t]
\setstretch{1.3} %
\centering
\caption{Quantitative results on multi-coil parallel MRI show that C-MSM can outperform diffusion-based solvers trained on clean priors in both PSNR and LPIPS.
\hlgreen{\textbf{Best values}} and \hlblue{second-best values} are color-coded for each metric.}
\small
\setlength{\tabcolsep}{4pt}
\renewcommand{\arraystretch}{1.00}
\vspace{2mm}
\resizebox{0.55\columnwidth}{!}{
\begin{tabular*}{0.6\columnwidth}{@{\extracolsep{\fill}}@{}clccc@{}}
\toprule
\textbf{Testing data} & \textbf{Method} & \textbf{PSNR} $\uparrow$ & \textbf{SSIM} $\uparrow$ & \textbf{LPIPS} $\downarrow$ \\
\midrule
\multirow{7}{*}{\textbf{CS-PMRI} $(\times 4)$}
& Input & 22.5 & 0.608 & 0.386 \\
& TV          & 25.9          & 0.738          & 0.320 \\
& A-DPS & 27.8 & 0.808 & 0.162 \\
& MSM         & 30.7          & 0.844          & 0.168 \\
& DPS         & 32.0           &  \hlblue{0.859}            & 0.162 \\
& DDNM        & \hlgreen{\textbf{32.9}}           & \hlgreen{\textbf{0.864}}            & \hlblue{0.159} \\
& C\text{-}MSM & \hlblue{32.7} & 0.853 & \hlgreen{\textbf{0.144}} \\ \addlinespace[2pt] \cdashline{1-5}  \addlinespace[2pt]
\multirow{7}{*}{\textbf{CS-PMRI} $(\times 8)$}
& Input & 21.5 & 0.541 & 0.445 \\
& TV          & 23.7          & 0.648          & 0.377 \\
& A-DPS & 25.6 & 0.742 & 0.302 \\
& MSM         & 27.6          & 0.751 & 0.287 \\
& DPS         & 27.9           & 0.742            & \hlgreen{\textbf{0.247}} \\
& DDNM        & \hlblue{28.6}           & \hlgreen{\textbf{0.764}}            & \hlblue{0.249} \\
& C\text{-}MSM & \hlgreen{\textbf{28.8}} & \hlblue{0.760} & 0.265 \\
\bottomrule
\end{tabular*}
}
\label{tab:recon_results}
\vspace{-.3cm}
\end{table*}

\begin{figure*}[t]
    \centering
    \includegraphics[width=1.0\linewidth]{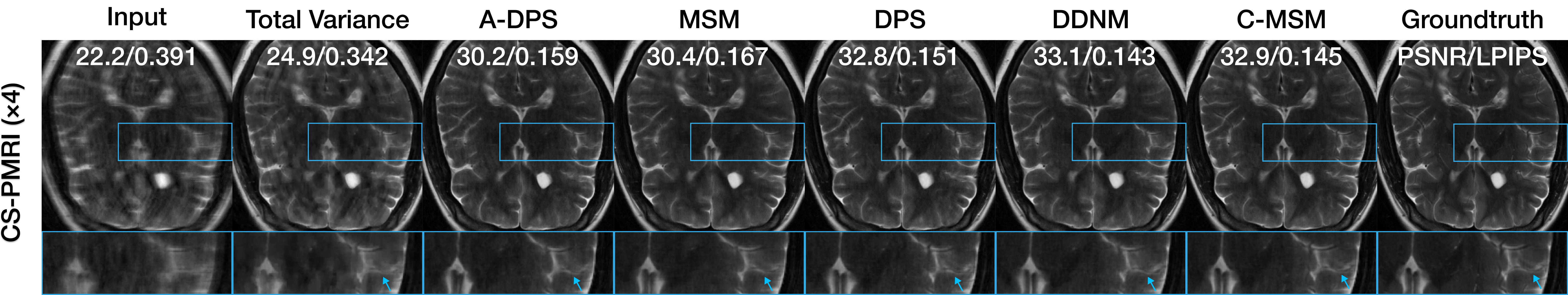}
    \rule[0pt]{1.0\linewidth}{0pt}
    \caption{Visual comparison of C-MSM with baseline methods on CS-PMRI under $\times4$ random sampling with noise level $\sigma = 0.01$. C-MSM performs close to DPS~\cite{chung2023dps} and DDNM~\cite{wang2023ddnm}, which rely on clean ground-truth images and pre-calibrated CSMs.}
    \label{fig:qual_grid}
\end{figure*}

\medskip\noindent
\textbf{Evaluation and Results.}
We compare C-MSM against both classical and diffusion-based baselines.
For the classical method, we include total variation (TV)~\cite{Block2007TV}, a regularized reconstruction method with an $\ell_1$ penalty tuned to $\lambda=0.002$ for the best validation peak signal-to-noise ratio (PSNR).
For clean prior-based diffusion-based baselines, DPS~\cite{chung2023dps} and DDNM~\cite{wang2023ddnm} are adapted as diffusion-based inverse problem solvers that use clean diffusion priors, with a step size of 1.7.
For self-supervised diffusion methods, MSM~\cite{park2025measurementscorebaseddiffusionmodel}, A-DPS~\cite{aali2025ambient}, and our C-MSM used the same step size of 2.0, while MSM and C-MSM used a stochastic mini-batch size of 10.

Reconstruction performance is measured using PSNR, structural similarity index measure (SSIM), and learned perceptual image patch similarity (LPIPS). 
Table~\ref{tab:recon_results} and Figure~\ref{fig:qual_grid} compare our self-supervised diffusion framework with automatic CSM estimation module to both self-supervised and supervised frameworks, all of which rely on pre-calibrated CSMs. The results show that jointly learning CSM estimation with the self-supervised denoising objective allows C-MSM to achieve performance comparable to the supervised framework.

\section{Conclusion}
We presented the C-MSM, a self-supervised diffusion framework for parallel MRI reconstruction that jointly estimates coil sensitivity maps (CSMs) and learns measurement scores directly from undersampled k-space data.
Experiments on the fastMRI dataset demonstrate that C-MSM consistently improves over MSM and achieves reconstruction quality close to clean prior-based solvers, despite pretrained without access to clean data and pre-calibrated CSMs. These results highlight the potential of C-MSM as a practical and calibration-free solution for diffusion-based MRI reconstruction.

\section{Acknowledgment}
This work was supported in part by the National Science Foundation under Grants No. 2504613 and No. 2043134 (CAREER).

{
\small

\bibliographystyle{IEEEbib}
\bibliography{references}
}



\end{document}